\documentclass[aps,preprint,amssymb,12pt,floatfix]{revtex4}
\usepackage{subfigure}
\usepackage{graphicx}
\usepackage{rotating}
\usepackage{tabularx}

\begin{document}

\title{Scenarios for protein aggregation: Molecular Dynamics simulations and Bioinformatic Analysis}

\renewcommand{\thefootnote}{\fnsymbol{footnote}}
\author{Ruxandra I. Dima$^{1}$, Bogdan Tarus$^{2}$, John E. Straub$^{2}$ and D. Thirumalai$^{3,4}$}
\affiliation{\em \small $^{1}$ Department of Chemistry, University of Cincinnati, Cincinnati, OH 45221\\
$^{2}$ Department of Chemistry, Boston University, Boston, MA 02215,\\
$^{3}$Biophysics Program, Institute for Physical Science and Technology\\
  $^{4}$Department of Chemistry and Biochemistry\\      
  University of Maryland, College Park, MD 20742}

\vspace{2cm}

\date{\small \today}

\baselineskip = 18pt

\begin{abstract}

\end{abstract}

\maketitle

\baselineskip = 22pt

{\bf Introduction:}

Increasing number of diseases including Alzheimer's disease
\cite{SelkoePhysiolRev01}, transmissible prion disorders
\cite{PrusinerPNAS98}, type II diabetes are linked to amyloid
fibrils \cite{ChitiAnnRevBiochem06}. The mechanism of amyloid fibril formation starting from the
monomer is still poorly understood. During the cascade of events in
the transition from monomers to mature fibrils a number of key
intermediates, namely, soluble oligomers and protofilaments are
populated. It is suspected that the conformations of the peptides in
these aggregated state differ substantially from the isolated monomer
which implies that the monomer undergoes substantial inter-peptide
interaction-driven structural transformations \cite{KooPNAS99}. The
need to understand the assembly kinetics of fibril formation has
become urgent because of the realization that soluble oligomers of
amyloidogenic peptides may be even more neurotoxic than the end
product, namely, the amyloid fibrils \cite{GlabeScience03}. In order
to fully understand the routes to fibril formation one has to
characterize the major species in the assembly pathways. The
characterization of the energetics and dynamics of oligomers (dimers,
trimers etc) is difficult using experiments alone because they undergo
large conformational fluctuations. In this context, carefully planned
molecular dynamics simulation studies
\cite{KlimovStruct03,CaflischPNAS03,TarusJMB05,BucheteJMB05},
computations using coarse-grained models \cite{DimaProtSci02}, and
bioinformatic analysis \cite{DimaBiophysJ02,DimaBioinfo04} have given
considerable insights into the early events in the route to fibril
formation. Here, we describe progress along this direction using
examples taken largely from our own work.

In this chapter, we focus on the following aspects of protein
aggregation using A$\beta$-peptides and prion proteins as examples.

$\bullet$ What are the plausible scenarios in the transition from
monomers to amyloid fibril formation?  

$\bullet$ What features of the amyloidogenic peptides control the
growth kinetics of fibrils? Although the assembly mechanism is complex
the overall growth kinetics is determined largely by the charge states
and hydrophobicity of the monomers. 

$\bullet$ Can sequence and structural analysis be used to predict
specific patterns that are likely to be susceptible to aggregation? By
exploiting the sequence profiles and structures of the cellular form
of prions, PrP$^{C}$, we uncover the regions that are likely to
trigger the large conformational changes in the transition from
PrP$^{C}$ to the scrapie form, PrP$^{Sc}$.

$\bullet$ Is there an organizational principle in oligomer and fibril
formation? The formation of morphologically similar aggregates by a
variety of proteins that are unrelated in sequence or structure
suggests that certain general principles may govern
fibrillization. However, the vastness of sequence space and the
heterogeneity of environmentally-dependent interactions make
deciphering the principles of aggregation difficult. Nevertheless, we
will argue that oligomers and higher order structures form in such a
way that the number of intra- and inter-molecular hydrophobic
interactions are maximized and electrostatic repulsions are
minimized. The latter implies that the motifs that minimize the number
of salt bridges are preferred.

For the issues raised above we formulate tentative ideas using
phenomenological arguments and atom molecular dynamics (MD)
simulations. Using a number of experimental observations and results
from computer simulations certain general principles of amyloid
formation seem to be emerging. There are a number of unresolved issues
that remain despite significant progress. A few of these are outlined
at the end of the paper.\\

\noindent
{\bf Scenarios for peptide-association:} 

\noindent
The molecular details of the cascade of events that lead to the
formation of amyloid fibrils remain unknown because the species along
the aggregation pathways are highly dynamic and metastable. Indeed,
AFM images of protofibrils show that they undergo shape fluctuations
which implies a heterogeneous population of species. A number of
experimental studies suggest that fibril formation exhibits all the
characteristics of a nucleation growth process.  It is suspected that
the formation of a critical nucleus is the rate determining step in
the fibril formation \cite{LansburyAnnRev97}.  Once the critical
nucleus, whose very nature might be both depend on sequence as well as
external conditions, forms the fibril formation process is essentially
downhill in free energy.  The nucleation characteristics manifests
themselves in the appearance of a lag phase in the fibril
formation. The lag phase disappears if a seed or a preformed nucleus
is present in the saturated peptide solution.  The seeded growth of
fibrils has also been observed in simple lattice and off-lattice
models of protofibril formation.  From this perspective the general
scenario for explaining aggregation kinetics is in place. However, the
details of the process including the dependence of the growth kinetics
on the specifics of the sequence are not fully understood.

\begin{figure}[ht]
\includegraphics[width=4.5in]{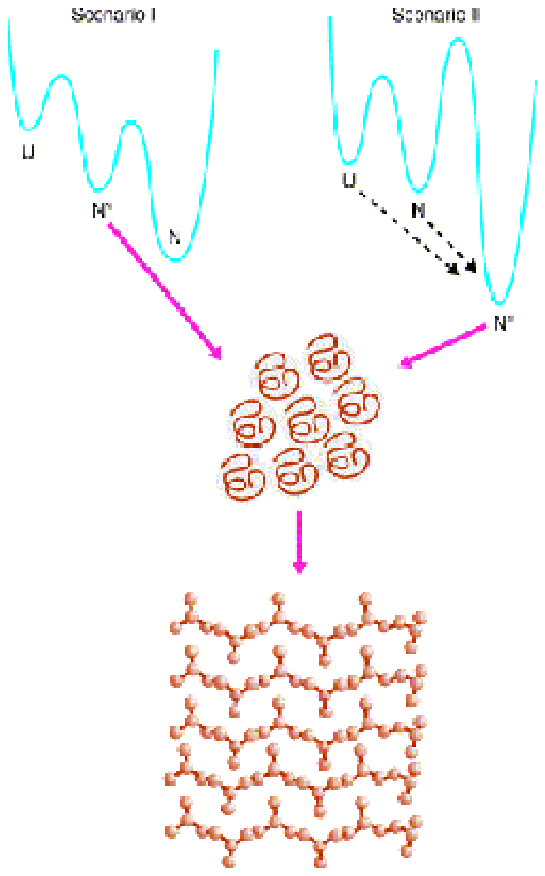}
\caption{\label{fig:Fig4_COSBreview}}
\end{figure}

Here we present two extreme scenarios \cite{ThirumalaiCOSB03} that
describe the needed conformational changes in monomers that lead to a
population of species that can nucleate and grow. The two potential
scenarios, which follow from the energy landscape perspective of
aggregation, differ greatly in the description of the dynamics of the
monomers. It was advocated early on that fibrillization requires
partial unfolding of the native state \cite{FinkFoldDes98} or partial
folding of the unfolded state (see Scenario I in
Fig. (\ref{fig:Fig4_COSBreview})).  Both events, which are likely to
involve crossing free energy barriers lead to the transient population
of the assembly-competent structures \textbf{N$^{*}$}. The better
appreciated possibility is Scenario I in which environmental
fluctuations (pH shifts for example) produces spontaneously the
\textbf{N$^{*}$} conformation. For example, extensive experiments
\cite{KellyCOSB98} have shown that the \textbf{N$^{*}$} state in
transthyretin (TTR), which has a higher free energy than the native
state \textbf{N}, formed upon unraveling of the strands C and D at the
edge of the structure. This process exposes an aggregation-prone
strand B. One can also envision a scenario in which \textbf{N$^{*}$}
has a lower free energy than \textbf{N} thus making the folded
(functional state) state metastable.  It is likely that amyloidogenic
proteins, in which nearly complete transformation the structure takes
place upon fibrillization, may follow the second scenario. In both
cases fibrillization kinetics results from the ability to populate the
\textbf{N$^{*}$} species. In either scenarios (TTR aggregation that
follows Scenario I or $PrP^{Sc}$ formation that follows scenario II)
growth kinetics is initially determined by the 'unfolding' barriers
separating \textbf{N$^{*}$} from either \textbf{N} or \textbf{U}. The
energy-landscape perspective for aggregation
(Fig. (\ref{fig:Fig4_COSBreview})) suggests that the free energy of
stability may not be a good indicator of fibril growth
kinetics. Rather, growth kinetics should correlate with unfolding
barriers.

In scenario I, the amyloidogenic state \textbf{N$^{*}$} is formed by
denaturation stress or other environmental fluctuations. The
production of {\bf N$^\star$} in scenario II can occur by two distinct
routes. If {\bf N} is metastable, as is apparently the case for
PrP$^c$ \cite{BaskakovJBC01}, then conformational fluctuations can
lead to {\bf N$^\star$}. Alternatively, formation of {\bf N$^\star$}
can also be triggered by intermolecular interactions. In the latter
case {\bf N$^\star$} can only form when the protein concentration
exceeds a threshold value. As noted below there is evidence for both
scenarios in the routes to fibril states.

In order to understand the kinetics of fibrillization it is necessary
to characterize the early events and pathways that lead to the
formation of the critical nucleus. In terms of the energy landscape,
the structures of {\bf N$^\star$}, the ensemble of transition state
structures, and the conformations of the critical nuclei must be known
to fully describe the assembly kinetics. Teplow and coworkers, who
have followed the growth of fibrils for eighteen peptides, including
A$\beta_{1-40}$ and A$\beta_{1-42}$ \cite{TeplowJMB01} showed that the
formation of amyloids is preceded by the transient population of the
intermediate oligomeric state with high $\alpha$-helical content. This
is remarkable given that both the monomers and fibrils have little or
no $\alpha$-helical content. Therefore, the transient formation of
$\alpha$-helical structure represent an on-pathway intermediate state.

Somewhat surprisingly, we found using multiple long MD simulations
that in the oligomerization of A$\beta_{16-22}$ peptides
\cite{KlimovStruct03} the oilgomer assembles into antiparallel
$\beta$-structure upon inter-peptide interactions. Even in the
oligomerization of these small peptides from the A$\beta$ family the
assembly was preceded by the formation of an on-pathway
$\alpha$-helical intermediate. Based on our findings and the work by
Teplow and coworkers we postulated that the formation of oligomers
rich in $\alpha$-helical structure may be a universal mechanism for
A$\beta$ peptides.

Formation of the on-pathway $\alpha$-helical intermediate may be
rationalized using the following arguments. The initial events involve
the formation of "non-specific" 
oligomers driven by  hydrophobic interactions that reduces the effective
available volume to each  A$\beta$ peptide. 
In the confined space peptides adopt  $\alpha$-helical
structure. Further structural changes are determined by the
requirement of maximizing the number of favorable  hydrophobic and
electrostatic interactions. Provided that A$\beta$ oligomers contains
large number of peptides, this can be achieved if A$\beta$ peptides 
adopt extended $\beta$-like conformations. 

There is some similarity between the aggregation mechanism postulated
for A$\beta$ peptides and the nucleated conformational conversion
(NCC) model envisioned for the conversion of Sup35 to [PSI$^+$] in
{\em Saccharomyces Cerevisiae} \cite{ChienNature01}. By studying the
assembly kinetics of Sup35, Serio {\em et al.}  \cite{Serio00Science}
proposed the NCC model, which combines parts of the templated assembly
and nucleation-growth mechanisms. The hallmark of the NCC model
\cite{Serio00Science} is the formation of a critical sized mobile
oligomer, in which Sup35 adopts a conformation that may be distinct
from its monomeric random coil or the one it adopts in the aggregated
state. The formation of a critical nucleus to which other Sup35 can
assemble involves a conformational change to states that it adopts in
the self-propagating [PSI$^+$]. The $\alpha$ helical intermediate seen
in A$\beta$ peptides may well correspond to the mobile oligomer that
has the "wrong" conformation to induce further assembly. \\

\noindent
{\bf Assembly of A$\beta_{16-22}$ oligomers:} 

\noindent
{\em The A$\beta_{16-22}$ monomer is a random coil:} The small size of
A$\beta_{16-22}$ peptides, which adopt an antiparallel $\beta$-sheet
structure in the fibril state, is an ideal system for exploring in
detail the mechanism of oligomer formation. Using fairly long and
multiple trajectories \cite{KlimovStruct03}, the assembly pathways for
3 A$\beta_{16-22}$ $\rightarrow$ (A$\beta_{16-22}$)$_{3}$ were probed
using all atom simulations in explicit water. The simulations of
A$\beta_{16-22}$ and the corresponding mutants allowed us to draw a
number of conclusions that may be of general validity.

The simulations of the A$\beta_{16-22}$ monomer at room temperature
and at neutral pH showed that it is predominantly a random coil. The
finite size of the system gives rise to large conformational
fluctuations that lead to the population of strand-like
structures. There is a very low ($\sim$ 3\%) probability of
$\alpha$-helical conformations. The study of this simple system shows
that the $\beta$-sheet conformation adopted by the monomer must be due
to interactions with other peptides.\\

\noindent
{\em Oligomerization of three A$\beta_{16-22}$ peptides requires a
transient monomeric $\alpha$-helical intermediate:} Upon interaction
with other peptides substantial changes in the conformations of the
individual monomers occur. The size of the monomer increases by about
50\%. More surprisingly, we found that as interpeptide interaction
increases there is a dramatic increase in the percentage of
$\alpha$-helical content during intermediate times. At longer times
the monomer undergoes a $\alpha \rightarrow \beta$ transition. Due to
the small size of the oligomer (n = 3) there are substantial
conformational fluctuations even after the three strands are roughly
in antiparallel registry. Nevertheless, the simulations showed that
the size of the nucleus for A$\beta_{16-22}$ cannot be large because
even with n = 3 there are signatures of stable oligomers. Indeed,
explicit simulations for t $>$ 300 ns show that one can obtain nearly
perfectly aligned A$\beta_{16-22}$ trimers in which the strands are in
antiparallel registry (Li, private communication). In these
simulations $< \hat{u}_i(t) \hat{u}_j(t) >$ where $\hat{u}_i(t)$ is
the unit vector connecting the N and C termini of peptide i fluctuates
around values close to -1.

The dominant pathway for 3 A$\beta_{16-22}$ $\rightarrow$
(A$\beta_{16-22}$)$_{3}$ from the simulations showed that in the
intermediate stages the monomer transiently populates an
$\alpha$-helix (see Fig.(8) from \cite{KlimovStruct03}). It should be
emphasized that in the assembly process (especially in the early
stages in the oligomerization) there are multiple routes. As a result,
kinetic trapping can result in structures that are not conducive to
forming the most stable antiparallel structures. Such kinetically
controlled structures have been explicitly probed in dimer formation
of small fragments of A$\beta$ peptides. These studies and other
simulations illustrate the complexity in dissecting the assembly of
even small amyloidogenic peptides into ordered structures.

The gross features of the fibril structures of a number of proteins
and peptides whose monomer sequences and structures are unrelated are
similar. This observation might suggest that the interactions that
stabilize the oligomers and fibrils must be ``universal'' involving
perhaps only backbone hydrogen bonds. It might appear that side
chains, and hence sequence differences, might play a secondary
role. Such a conclusion is further supported by repeated observations
\cite{ChitiPNAS99} that any protein or peptide can be made to form
cross-$\beta$ structures under appropriate conditions. However,
experiments \cite{IvanovaPNAS04} and simulations
\cite{KlimovPNAS04,CaflischPNAS03} show that side chain interactions are
crucial in directing oligomer formation. Trimers of A$\beta_{16-22}$
are stabilized primarily by favorable inter-peptide hydrophobic
interactions between residues in the central hydrophobic cluster
(LVFFA) and secondary by inter-peptide salt bridge formation between K
and E.

The importance of side chains can be demonstrated by examining the
effect of mutations on the trimer formation in A$\beta_{16-22}$
peptides. We showed using simulations that the mutant GLVFFAK, which
eliminates the formation of intermolecular salt-bridge entirely
destabilizes the trimer. Similarly replacement of L, F, F by S, also
destabilizes the trimer. These simulations show that the sequence
plays a key role in the tendency of peptides to form amyloid
fibrils. Although no general role has emerged it seems that sequences
with enhanced correlation between charges \cite{DimaBioinfo04} or
preponderance of contiguous ($>$ 3) hydrophobic residues might be
amenable to amyloid formation on finite time scales.\\

\noindent
{\bf Dimerization of A$\beta_{10-35}$ peptides:} 

\noindent
{\em Generation of putative dimer structures:} In contrast to
A$\beta_{16-22}$ fibrils the longer peptide A$\beta_{10-35}$ adopts a
parallel $\beta$-sheet conformation in the amyloid state. It is now
suspected that in the fibril state the monomer is stabilized by an
intra-molecular salt bridge between Asp23 and Lys28. In order for this
salt bridge to form there has to be a bend in the monomeric structures
involving the residues VGSN. The importance of a stable turn, which
was experimentally determined in the NMR structures, was emphasized in
MD simulations as well.

\begin{figure}[ht]
\subfigure[]{
	  \label{fig:Fig2a_JMBTarus}
          \includegraphics[width=2.75in]{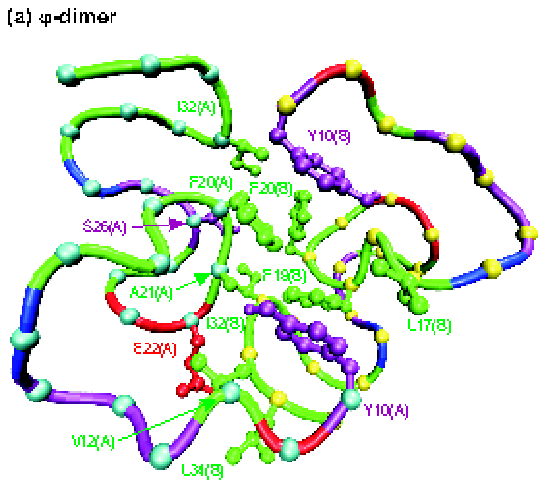}}\\
        \subfigure[]{
        \label{fig:Fig2b_JMBTarus}
        \includegraphics[width=2.75in]{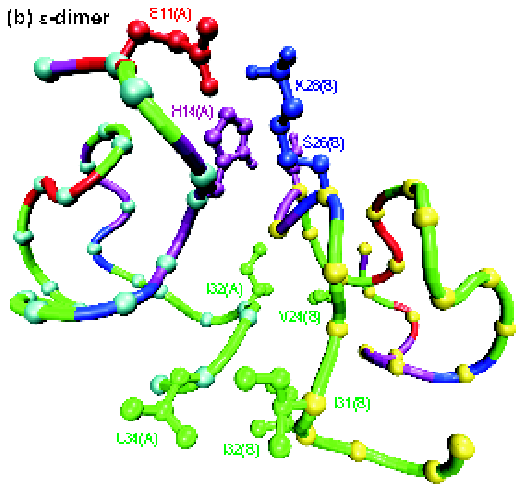}}
\caption{\label{fig:Fig2_JMBTarus}}
\end{figure}

In a recent study \cite{TarusJMB05}, we used a number of computational
methods to probe dimer formation. We first generated a putative set of
dimer conformations that is based on shape complementarity. The work
on A$\beta_{16-22}$ showed that both inter-peptide hydrophobic
interactions and the creation of favorable electrostatic contacts are
required to produce marginally stable oligomers. In order to dissect
their relative importance we generated two homodimer decoy sets by
maximizing the number of contacts between the monomer interfaces. The
first 2000 dimer structures of each set were selected by minimizing
the interaction energy between the monomers. In order to distinguish
between desolvation and electrostatic interactions we used two
distinct energy functions. The \mbox{$\varphi$-dimer}
(Fig. (\ref{fig:Fig2a_JMBTarus})) minimizes the desolvation energies
of the dimer at the interface whereas the \mbox{$\varepsilon$-dimer}
(Fig. (\ref{fig:Fig2b_JMBTarus})) corresponds to structures that have
the highest inter-peptide electrostatic interactions. The structure of
the \mbox{$\varphi$-dimer} is dominated by contacts between
hydrophobic segments of the monomers. The hydrophobic core,
LVFFA(17-21), and the hydrophobic C-terminus of both monomers are
buried at the dimer interface. The contacts at the interface of the
\mbox{$\varphi$-dimer} are conserved over the lowest energy dimer
structures. The \mbox{$\varepsilon$-dimer} interface is characterized
by electrostatic inter-monomeric interactions, among which the salt
bridge Glu11(A)-Lys28(B) has the largest contribution. Contrary to the
\mbox{$\varphi$-dimer}, the contacts observed at the
\mbox{$\varepsilon$-dimer} interface are not conserved across the set
of the low energy dimers due to the increased specificity and strength
of the electrostatic interaction.\\

\begin{figure}[ht]
\includegraphics[width=6.0in]{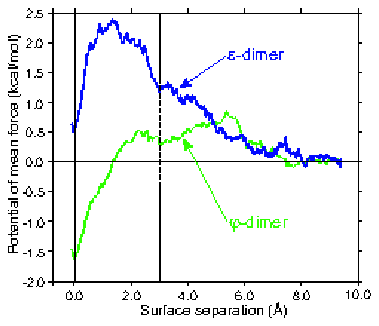}
\caption{\label{fig:Fig5_JMBTarus}}
\end{figure}

\noindent
{\em Interior of A$\beta$ oligomers is dry:} Insights into the
assembly mechanism of the \mbox{$\varphi$-dimer} and
\mbox{$\varepsilon$-dimer} can be obtained from the Potential of Mean
Force (PMF). The PMF for the dimerization process was obtained along
the center of mass of the two monomers as the
Fig.(\ref{fig:Fig5_JMBTarus}) indicates. For each free energy profile,
one can distinguish three distinct intervals. In the outer interval,
the PMF value is nearly constant, from 6.5~\AA\ -- 7.0~\AA\ to maximum
separation, which in our case is 9.0~\AA. At a distance of 6.5~\AA\
for the \mbox{$\varepsilon$-dimer} and 7.0~\AA\ for the
\mbox{$\varphi$-dimer}, the first solvation shells of the monomers
come into contact, and for both dimers the energetics of desolvation
of the associating monomers is unfavorable. In the second interval for
the \mbox{$\varepsilon$-dimer}, the value of the PMF continues to
increase up to 1.2~kcal/mol at a 3.0~\AA\ separation; for the
\mbox{$\varphi$-dimer}, the potential energy reaches a value of
0.8~kcal/mol at 5.5~\AA, and after that the desolvation is favorable,
ending in an unstable local minimum at 3.0~\AA. For the third
interval, from 3.0~\AA\ to 0.0~\AA, there is only one solvation shell
between the monomers. The water molecules are most strongly ordered
near the monomers through electrostatic interactions and hydrogen
bonds. As a result, the PMF for the \mbox{$\varepsilon$-dimer}
increases sharply between 3.0~\AA\ and 1.3~\AA\ up to 2.4~kcal/mol. At
contact, the van der Waals attraction predominates, making the overall
dimerization process energetically favorable. For the
\mbox{$\varphi$-dimer}, the solvation shell between the hydrophobic
regions of the monomers is only weakly bound to the solute, and after
a small increase in the PMF, corresponding to the van der Waals
attraction, the desolvation is entirely favorable.

If the approach along the center of mass of the monomers approximately
represents a minimum energy path, then the expulsion of water in the
\mbox{$\varphi$-dimer} must be an early event in the
assembly. Explicit simulations for A$\beta_{16-22}$ oligomers
\cite{KlimovStruct03} also show that desolvation occurs early. As a
result, the interior of A$\beta$ oligomers is dry.\\

\noindent
{\em Hydrophobic interactions between monomers are the driving force
in the association of A$\beta_{10-35}$ peptides into dimers:}
Comparing the \mbox{$\varphi$-dimer} \mbox{$\varepsilon$-dimer} models
for monomer association, we find that the former appears to lead to
more energetically favorable dimerization than the latter. It appears
to be more efficient to remove the entropically unfavorable structured
water between the opposing hydrophobic regions of the two monomers
than to stabilize the monomer solely through electrostatic
interactions. This is in good agreement with the experimental and MD
simulations observation that the mutation E22Q -- where a charged
glutamic acid residue is replaced by a polar glutamine residue --
increases the propensity for amyloid
formation\cite{Miravalle2000,Esler2000}. Molecular dynamics
simulations of this increased amyloidogenic activity for the E22Q
mutant peptide led to the conclusion that the water-peptide
interaction is less favorable for the mutant peptide
\cite{Massi2001b}. Following a more detailed analysis of the structure
and dynamics of the WT and E22Q A$\beta_{10-35}$, it has suggested
that a change in the charge state of the peptide, due to the E22Q
mutation, leads to an increase of the hydrophobicity of the peptide
that could be responsible for the increased activity\cite{Massi2002}.

The time evolution of the \mbox{$\varphi$-dimer} structure was
analyzed and it was observed that the monomers remain in contact
during the simulation. It was shown that the hydrophobic interaction
between the monomers of the \mbox{$\varphi$-dimer} acts as a
stabilizing force of the dimer. The ``extended core'' region 15--30 of
both monomers in the \mbox{$\varphi$-dimer} makes the principal
contribution to the hydrophobic interaction energy. The
\mbox{$\varphi$-dimer} undergoes internal structural reorganization in
the terminal regions of the monomeric peptides. Our simulations
indicate that there is substantial reorganization of the peptide
monomers in the N- and \mbox{C-terminus} regions, as expected for a
dimer weakly and relatively non-specifically stabilized by hydrophobic
contacts at the dimer interface. Importantly, the structure of the
central hydrophobic cluster LVFFA region assumes a conformation
similar to that observed for the monomeric peptide in both
experiment\cite{Zhang1998} and simulation\cite{Massi2001a}. Our
simulations suggest that the preservation of the structure of the
LVFFA central hydrophobic cluster plays an important role in the
stabilization of the \mbox{$\varphi$-dimer} structure.

The finding that the \mbox{$\varphi$-dimer} may constitute the
ensemble of stable A$\beta_{10-35}$ dimer has important implications
for fibril formation. The initial event in the dimerization involves,
in all likelihood, contacts between the central hydrophobic
clusters. In this process, expulsion of water molecules in the
interface might be a key event just as in the oligomerization of
A$\beta_{16-22}$ fragments\cite{KlimovStruct03}. Since this process
involves cooperative rearrangement of ordered water molecules, it is
limited by an effective free energy barrier. Based on our results, we
conjecture that events prior to the nucleation process themselves
might involve crossing free energy barriers which depend on the
peptide-peptide and peptide-water interactions
(Fig. (\ref{fig:Fig4_COSBreview})).\\

\noindent
{\bf Initial stages in the PrP$^{C}$ conformational transition:}

\begin{figure}[ht]
\includegraphics[width=6.0in]{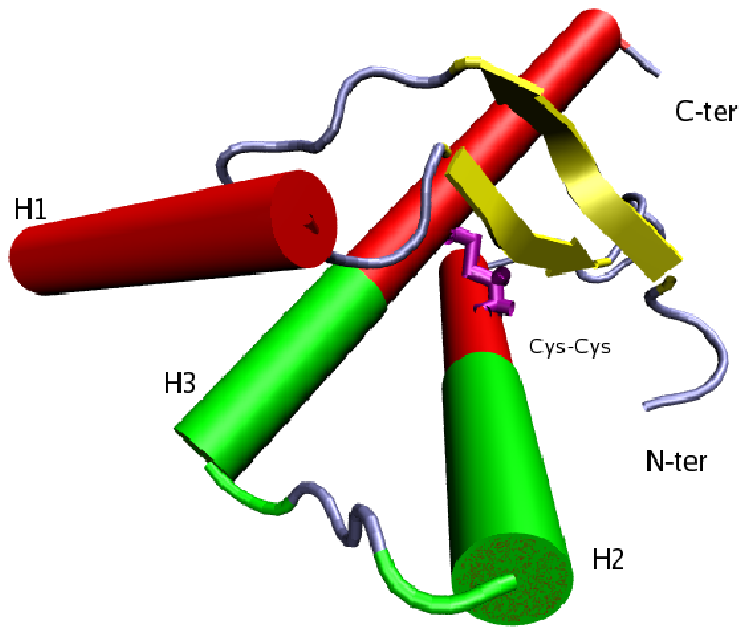}
\caption{\label{fig:1qlx}}
\end{figure}

\noindent
Prion proteins are extracellular globular proteins which are attached
to the plasma membrane by a GPI anchor. They have been linked to
various transmissible spongiform encephalopathies (TSEs) including the
bovine spongiform encephalopathy, the scrapie disease in sheep, and
the Creutzfeldt-Jakob disease in humans. The causative agent in these
diseases is believed to be the aggregated form (PrP$^{Sc}$) of the
cellular prion protein (PrP$^C$) \cite{RiekNature96}. The transition
to the scrapie form involves a large conformational change from the
mainly $\alpha$-helical PrP$^{C}$ to the PrP$^{Sc}$ state that is rich
in $\beta$-sheet.  According to the ``protein-only'' hypothesis
\cite{PrusinerPNAS98} PrP$^{Sc}$ serves as a template in inducing
conformational transitions in PrP$^{C}$ that can subsequently be added
to PrP$^{Sc}$.  The ``protein-only'' hypothesis implies that the
conformational change leading to the PrP$^{Sc}$ formation from the
normal cellular form PrP$^{C}$ may be spontaneous or might involve
interactions with unidentified protein X \cite{TellingCell95}. Prion
proteins, encoded by a single gene, consist of about 250 residues of
which the first 22 form a signal sequence. This is followed by
unstructured, but likely helical, Cu$^{2+}$ binding octarepeats rich
in glycine \cite{PrusinerPNAS98}. The NMR
\cite{RiekNature96,DonnePNAS97,ZahnPNAS00} and X-ray \cite{HaireJMB04}
structure of PrP$^{C}$ in various species (human, mouse, syrian
hamster, bovine, and sheep) shows that the ordered C-terminal part is
composed of a short antiparallel $\beta$-sheet that contains 8\% of
the residues in the (90-231) fragment and three helices representing
48\% of the secondary structure (Fig.(\ref{fig:1qlx}). Fourier transform infrared
spectroscopy measurements \cite{CaugheyBiochem91,GassetPNAS93}
indicate that PrP$^{Sc}$(90-231) has 47\% $\beta$-sheet and 24\%
$\alpha$-helical content.

We have suggested using structural, bioinformatic, and molecular
dynamics simulations that formation of PrP$^{Sc}$ follows scenario II
(see Fig. (\ref{fig:Fig4_COSBreview})). This implies that, either
spontaneously or in the presence of a seed of PrP$^{Sc}$, the
metastable cellular form, PrP$^{C}$, undergoes a transition to the
PrP$^{C*}$ state that is capable of further aggregating or adding to
an already present PrP$^{Sc}$ particle. Experiments
\cite{KuwataBiochem02} and scenarios of protein aggregation
\cite{ThirumalaiCOSB03} suggest the proposal that the conformational
transition involves the formation of PrP$^{C*}$ that is more stable
than PrP$^{C}$. The transition from the metastable PrP$^{C}
\rightarrow$ PrP$^{C*}$, which involves crossing a substantial free
energy barrier on the order of 20 kcal/mole
\cite{BaskakovJBC01,BaskakovJBC02}, results in a state that can
nucleate and polymerize to the protease resistant form.  We also
identified the putative regions that are involved in the $PrP^{C}
\rightarrow PrP^{C*}$ transition.  Comparison of a number of
structural characteristics (such as solvent accessible area,
distribution of ($\Phi,\Psi$) angles, mismatches in hydrogen bonds,
nature of residues in local and non-local contacts, distribution of
regular densities of amino acids, clustering of hydrophobic and
hydrophilic residues in helices) between PrP$^C$ structures and a
databank of "normal" proteins shows that the most unusual features are
found in helix 2 (residues 172-194) followed by helix 1 (residues
144-153) \cite{DimaBiophysJ02}. In particular, the C-terminal residues
in H2 are frustrated in their helical state.  Application of the
recently introduced notion of discordance, namely, incompatibility of
the predicted and observed secondary structures, also points to the
frustration of H2 not only in the wild type but also in mutants of
human PrP$^C$. This suggests that the instability of PrP$^C$ proteins
may play a role in their being susceptible to the profound
conformational change.

We showed \cite{DimaBiophysJ02} that, in addition to the previously
proposed role for the segment (90-120) and possibly H1, the C-terminus
of H2 and possibly N-terminus may play a role in the $\alpha
\rightarrow \beta$ transition. Sequence alignments show that helices
in avian prion proteins (chicken, duck, crane) are better accommodated
in a helical state which might explain the absence of PrP$^{Sc}$
formation over finite time scales in these species. From the analysis
it is clear that the conformational fluctuations in the C-terminal end
of helix 2 (H2) and in parts of helix 3 (H3) are involved in the
transition to PrP$^{C*}$. Because the stability of PrP$^{C}$ arises
from the structures in the C-terminal end, the transition to
PrP$^{C*}$ requires global unfolding of PrP$^{C}$ \cite{HosszuNSB99}
which explains the origin of the high free energy barrier separating
PrP$^{C}$ and PrP$^{C*}$ \cite{DimaBiophysJ02}. NMR experiments
\cite{KuwataBiochem02,KuwataBiochem04} showed that conformational
fluctuations that originate in the C-terminal part of H2 are essential
in the formation of PrP$^{C*}$. Structural and mutational studies have
also shown that the relatively short helix 1 (H1) is stable over a
range of pH values and solvent conditions, and hence is unlikely to
undergo conformational change in the transition to PrP$^{C*}$
\cite{LiuBiopolym99,SpeareJBC03,ZieglerJBC03}.

The required conformational fluctuations in PrP$^{C}$ needed to
populate PrP$^{C*}$ suggest that the earliest event involves extensive
unfolding of the monomeric PrP$^{C}$. We used results from a database
search of sequence patterns in helices of PrP$^{C}$ and extensive all
atom molecular dynamics (MD) simulations of helical fragments from the
mouse prion protein (mPrP$^{C}$) to shed light on the nature of
instabilities that drive the PrP$^{C} \rightarrow$ PrP$^{C*}$
transition \cite{DimaPNAS04}.  Previously MD simulations have been
used to probe other structural aspects of prion proteins including
structures of protofibrils \cite{DeMarcoPNAS04}. The 10 residue H1,
with an unusual sequence pattern (highly charged and presenting the
largest percentage of salt bridges in any $\alpha$-helix in the PDB),
remains helical for the duration of the simulation ($\approx$ 0.09
$\mu$s). The double mutant (D147A,R151A), which eliminates one of the
three salt bridges in H1, is less stable than the wild type. Multiple
MD trajectories of peptides encompassing H2 and H3 (together with
their connecting loop) with intact disulfide bond (Cys179-Cys214)
showed that residues in the second half of H2 clustered around
positions 187-188 have large conformational flexibility and non-zero
preference for $\beta$-strand or coil-like structures. Instability in
H2 propagates to H3 especially from position 214 onwards. Based on
these results, we mapped the plausible structures of the aggregation
prone PrP$^{C*}$.  Despite the limitations (short simulation time and
the expected variations of results with different force fields) of all
atom simulations, different computational approaches yield
qualitatively similar results, namely, the initial conformational
transition must involve at least partial unfolding of parts of H2 and
H3.\\

\noindent
{\bf Bioinformatic Analysis:}

\noindent
{\em Pattern of charges in H1 is rare: } The distribution of $R(+,-)$
for the 2103 helices from the DSMP shows
that {\em no other natural sequence} has as many (+,-) pairs at
positions (i,i+4) as H1 from PrP$^C$.  The search of the entire
PDBselect database for the H1 charge pattern shows that in only 56
(4.6\%) sequences this pattern occurs at least once, with the total
number of patterns being 63.  If we restrict the search to be the
exact pattern of H1, i.e. i = - , i+3 = -, i+4 = +, i+7 = + and i+8 =
- the number of sequences is a mere 23 (or 1.9\%).  Ziegler et
al. \cite{ZieglerJBC03} arrived at a similar conclusion based on a H1
pattern search in PDB.  The 23 rare sequence fragments are either
$\alpha$-helical (83\%) or in a random coil state (17\%).  Analysis of
the Yeast genome shows that 828 (or 9.2\%) of sequences have the
general pattern of H1 with only 253 (2.8\%) having the exact pattern.
In the {\it E. Coli} genome the numbers are: 158 (3.7\%) for the
general charge pattern and 51 (1.2\%) for the exact match.  These
results suggest that the sequence of H1 in PrP$^{C}$ is unusual not
only in its high charge content, but also in the {\em positioning of
charges along the sequence}.  More importantly, for the 23 proteins
with known 3D structures, the exact charge pattern results
overwhelmingly in $\alpha$-helices. Even more interestingly, analysis
of the 19 sequences with mostly $\alpha$-helical structure reveals
that the majority (88\%) of (+,-) pairs of residues found at positions
(i,i+4) form salt bridges.  These results indicate that the unusual
stability of the short helix H1 is possibly associated with its
ability to form the highly stabilizing salt bridges involving (i,i+4)
residues.\\

\noindent
{\em Pattern of hydrophobicity in H2 is rare:} There are very few
sequences that share the pattern of hydrophobicity of H2. In
PDBAstral40 \cite{ChandoniaNAR02} (proteins in the PDB having at most
40\% sequence similarity) there are only 12 (0.2\%) such sequences. In
the {\em E Coli} genome the number is 46 (1\%), while in the Yeast
genome it is 122 (1.4\%).  Inspection of the structures of the 12
proteins from PDBAstral40, shows that the sequence is never entirely
helical! For example, in only 13\% of these proteins the last 5
residues are found in a helix.  A characteristic pattern seen in H2
from mammalian prion proteins is TTTT (positions 190-193).  In the
PDBAstral40 this pattern occurs in only 18 proteins, including the
prion sequence. In an overwhelming number of these cases (15 of the 18
proteins) the TTTT pattern is found in a strand and/or loop
conformation (irrespective of the identity of the flanking amino
acids).  These results add further support to our proposal
\cite{DimaBiophysJ02} that the second half of H2 would be better
accommodated in non-helical conformations. \\

\noindent
{\em "Frustrated" secondary structural elements may be  harbinger of tendency to polymerize }
The ease of aggregation and the morphology of the aggregates depend
not only on the protein concentration, but also on other external
conditions such as temperature, pH and salt concentration. Although
most proteins can, under suitable conditions, aggregate the
observation that several disease causing proteins form amyloid fibrils
under physiologically relevant conditions raises the question: Is
aggregation or the need to avoid unproductive pathways encoded in the
primary sequence itself ? It is clear that sequences that contain a
patch of hydrophobic residues are prone to form aggregates
\cite{WestPNAS99}. However, it is known that contiguous patches (three
or more hydrophobic residues) occur with low probability in globular
proteins \cite{SchwartzProtSci01}. For example, sequences with five
hydrophobic residues (LVFFA in A$\beta$ peptide) in a row are not well
represented. Similarly, it is unusual to find hydrophobic residues
concentrated in a specific region of helices such as is found in helix
2 in PrP$^C$ \cite{DimaBiophysJ02}. 

It is natural to wonder if secondary structure elements bear
signatures that could reveal amyloidogenic tendencies. Two studies have
proposed that the extent of ``frustration'' in the secondary structure
elements (SSE) may be harbinger of amyloid fibril formation
\cite{DimaBiophysJ02,KallbergJBC01}. Because reliable secondary
structure prediction requires knowing the context dependent
propensities and multiple sequence alignments (such as used in PHD,
Profile network from Heidelberg \cite{RostJMB93}), it is more likely
that assessing the extent of frustration in the SSE rather than
analysis of sequence patterns is a better predictor of fibril
formation. Frustration in SSE is defined as the incompatibility of the
predicted (from PHD, for example) secondary structure and the
experimentally determined structure \cite{KallbergJBC01}.  For
example, if a secondary structure is predicted to be in a
$\beta$-strand with high confidence and if that segment is found (by
NMR or X-ray crystallography) to be in a helix, then the structure is
frustrated (or discordant or mismatched). The $\alpha$/$\beta$
discordance, which can be correlated with amyloid formation, can be
assessed using the score $S_{\alpha/\beta} = \frac{1}{L}
\sum_{i=1}^{L} (R_i - 5)$, where $R_i$ is the reliability score
predicted by PHD at position $i$ of the query sequence, 5 is the mean
score, and $L$ is the sequence length. The bounds on
$S_{\alpha/\beta}$ are $0 \leq S_{\alpha/\beta} \leq 4$ with maximal
frustration corresponding to $S_{\alpha/\beta}$ = 4. Similarly, the
measure $S_{\beta/\alpha}$ gives the extent of frustration of a
stretch that is predicted to be helical and is found experimentally to
be a strand. Using $S_{\alpha/\beta}$ and other structural
characteristics, one can make predictions of the plausible regions
that are most susceptible to large conformational fluctuations.\\

\noindent
{\it PrP$^{C}$ and Dpl}: Using the above concept of SSE frustration
the 23 residue sequence (QNNFVHDCVNITIKQHTVTTTTK) in mouse PrP$^C$,
with a score of 1.83, was assessed to be frustrated or discordant
\cite{DimaBiophysJ02}. Other measures of quantifying the structure
also showed that the maximal frustration is localized in the second half
(C-terminal of H2) \cite{DimaBiophysJ02}. The validity of this
prediction finds support in the analysis of mutants of the PRNP gene
associated with inherited TSEs (familial CJD, and FFI). According to
SWISS-PROT \cite{BairochNAR00} seven disease-causing point mutations
(D178N, V180I, T183A, H187R, T188R, T188K, T188A) are localized in
H2. We have used the sequence numbering for the mPrP$^C$. A naive use
of propensities to form helices, a la Chou-Fasman \cite{ChouARB78},
would suggest that with the exception of D178N all other point
mutations should lead to better helix formation. However, the
$S_{\alpha/\beta}$ scores for the mutants are 1.94, 1.80, 1.30, 1.80,
1.54, 1.94, and 1.94 for D178N, V180I, T183A, H187R, T188K, T188R, and
T188A respectively. Thus, in all these mutants H2 is frustrated making
it susceptible to conformational fluctuations which have to occur
prior to fibrillization. The differences in $S_{\alpha/\beta}$, which
can be correlated with local stability, suggest that stability alone
might not be a good indicator of the kinetics of amyloid formation.

\begin{figure}[ht]
\includegraphics[width=6.0in]{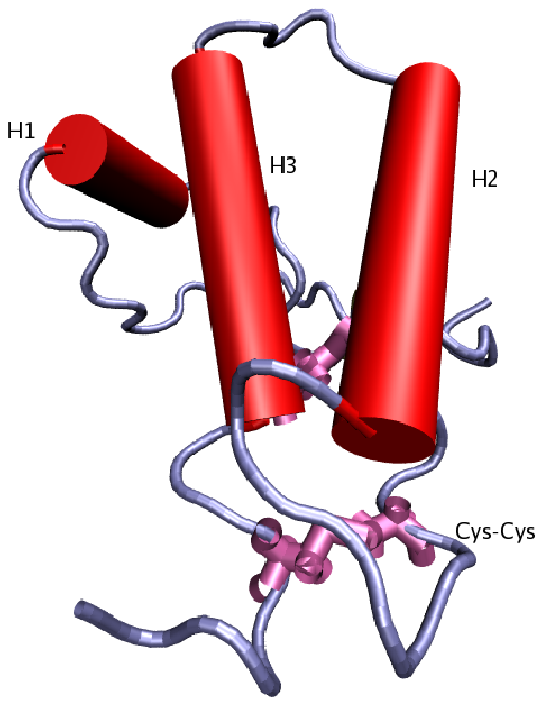}
\caption{\label{fig:1lg4}}
\end{figure}

The gene coding for the Doppel protein (Dpl), termed $Prnd$
\cite{MooreJMB99}, is a paralog of the prion protein gene, $Prnp$, to
which it has about 25\% identity. Normally, Dpl is not expressed in
the central nervous system, but it is up-regulated in mice with
knockout $Prnp$ gene.  In such cases, overexpression of Dpl causes
ataxia with Purkinje cell degeneration \cite{MooreJMB99}, which in
turn can be cured by the introduction of one copy of wild-type PrP
mouse gene \cite{NishidaLabInvest99}. NMR studies of the
three-dimensional (3D) structure of mouse Dpl \cite{MoPNAS01} showed
that it is structurally similar to the structure of PrP$^C$
(Fig. (\ref{fig:1lg4})).  However, PrP$^C$ and Dpl produce
diseases of the central nervous system using very different
mechanisms: PrP$^C$ causes disease only after conversion to the
PrP$^{Sc}$ form, while simple overexpression of Dpl, with no necessity
to form the scrapie form, causes ataxia.  The markedly different
disease mechanisms of PrP and Dpl would suggest, in light of the
findings for PrP$^C$, that the mouse Dpl (PDB code 1i17) would not be
frustrated.  Indeed, prediction of secondary structure by PHD
\cite{RostJMB93} on mouse Dpl correlates well with the experimentally
derived structure.  The only difference between the predicted and the
derived structure in Dpl is found in the first $\beta$-strand region
which is predicted to be helical by PHD. But the corresponding
$S_{\beta/\alpha}$ = -3.0 indicating that this $\alpha$-helix
prediction is unreliable as this sequence has low complexity. Also,
the analysis of 1i17 with the WHAT CHECK program \cite{HooftNature96}
reveals that, on average, there are only 8 unsatisfied buried
hydrogen-bond donors/acceptors representing 7.4\% of all residues in
mouse Dpl. This is comparable with the average value of 6\% found in
normal proteins, but it is quite smaller than the 14\% value seen in
mPrP (PDB code 1ag2). This analysis rationalizes the lack of
observation of scrapie formation in Dpl.\\

\noindent
{\bf Molecular Dynamics simulations:}

\noindent
{\em Helix 1 in mPrP is stable:} In order to dissect the stability of
PrP$^{C}$ fragments that were identified using bioinformatic analysis,
we used MD simulations of H1 and H2 and H3 from the PrP$^{C}$ state.
With the exception of residues 150-152, the propensities of the
interior residues for $\alpha$-helical or $\beta$-strand conformations
show that the helical structure is overwhelmingly preferred.  The
distribution of distances between residues at positions (i,i+4)
averaged over the 5 trajectories shows that, with the exception of
residues in the second half of H1, the helical structure is
preserved. Snapshots of typical conformations at various moments along
one of the trajectories show that even the C-term end of H1, which
becomes disordered after $\sim$ 12 ns, returns to the helical
conformation towards the end of the run.  Small fluctuations in a
short helix are unusual because it is known that isolated helices are
at best marginally stable \cite{BundiBiopolym79,DysonNSB98}.\\

In order to check if the predicted stability of H1 depends on the force field
we generated 2 trajectories for a total of 40 ns using the Charmm27
parameter set with the package NAMD. The backbone rmsd with respect to
the PDB structure stabilizes around 2.5-3.0 \AA\ after $\sim$ 10
ns. The rmsd for the backbone of the 144-149 fragment of the chain
remains close to 0.5 \AA\ for the duration of the run which is in very
good agreement with the previous set of simulations. The difference
between these two sets of simulations is only in the fraying of the
C-terminus residues. These results, which are consistent with the MOIL
simulations, also show that the fraction of helix content in H1 is
high. \\

\noindent
{\em Mutations of residues in the second salt bridge (D147-R151)
enhance conformational fluctuations:} The pattern searches suggest
that the three (i,i+4) salt bridges ((Asp144,Arg148), (Asp147,Arg151),
and (Arg148,Glu152)) in H1 should stabilize the isolated H1.  To probe
the importance of the second salt bridge (Asp147,Arg151), we simulated
the double mutant H1[D147A,R151A]. Replacing D and R by A should not
compromise the local helical propensity because Ala is the best
helix-former among the amino acids \cite{Creightonbook93}.
Consequently, any loss of stability in the structure can be attributed
largely to the loss of the the salt bridge. From relatively long MD
simulations for H1[D147A,R151A], we find that the double mutant has
increased conformational flexibility compared to the wild-type chain.
Most residues, except position 145, have non-zero $\beta$-strand
propensity.

The larger conformational fluctuations result in extended states with
only the first turn of the helix still present.  Based on these
findings, we conclude that H1[D147A,R151A] populates two basins of
attraction: one that is predominantly $\alpha$-helical with a radius
of gyration $\sim$ 6 \AA, and the other being mostly RC with a radius
of gyration of $\sim$ 7.7 \AA.  Time evolution of distances between
(i,i+4) residues (data not shown) shows that the conformational change
starts towards the C-terminus part of the sequence and proceeds in a
highly cooperative manner.  Our findings are in agreement with recent
experiments \cite{ZieglerJBC03} which showed that the peptide
huPrP(140-158)D147A is destabilized compared to wt-huPrP(140-158). The
decreased stability of the mutant could result in the efficient
conversion of PrP$^{C}$(90-231) to the protease resistant form.

By classifying the structures generated in the MD simulations as
helical \cite{KlimovStruct03}, we find that the helical fraction,
$f_{H}$, of the mutant is 0.55 while $f_{H}$ for the WT is 0.64. The
value of $f_{H}$ is 0.63 using the Charmm parameters.  We should {\em
emphasize} that the absolute values of $f_{H}$ might be overestimated
and could depend upon the force field. However, meaningful conclusions
can be drawn using the relative values. Using the $f_{H}$ values we
can estimate the free energy of stability using $\Delta F = - RT
ln(\frac{f_{H}}{1 - f_{H}})$. For the WT $\Delta F_{WT} \sim$ -0.37
kcal/mol, whereas for H1[D147A,R151A] $\Delta F_{M} \sim$ -0.13
kcal/mol.  If $f_H$ using the Charmm parameter set is used, then
$\Delta F_{WT} \sim$ -0.34 kcal/mol.  The relative difference $\Delta
\Delta F = \Delta F_{WT} - \Delta F_{M} \sim$ -0.24 kcal/mol which
arises from the salt bridge formation in WT. Interestingly, this
estimate for free-energy gain due to salt bridge formation is in the
range of the values reported in the literature
\cite{MakhatadzeJMB03}. \\

\noindent
{\em Second half of H2 is susceptible to conformational fluctuations:}
The trajectories, obtained using the NAMD package for a total of 185
ns, showed a drastic reduction in the amount of helical structure
which is accompanied by an increase in $\beta$-strand content.  The
conformational transition starts in the second half of H2 and
propagates towards its N-terminal, while H3 unwinds concomitantly at
its two ends.  The propensities of residues for $\alpha$-helical or
$\beta$-strand conformations show that only positions 178 and 179 (H2)
and residues 205 to 212 (H3) maintain their native $\alpha$-helical
structure. The extent of the conformational transition is also
reflected in the behavior of the backbone rmsd from the PDB structure
(1ag2) which increases monotonically from 3 \AA\ to 6 \AA\ in about 5
ns and reaches 11 \AA\ in the next 70 ns.

\begin{sidewaysfigure}[htbp]
  \vspace{0.1cm}
        \hspace{-2.0in}
        \subfigure[]{
	  \label{fig:conf_H23mPrP_SSon_Namd_init}
          \includegraphics[width=2.75in]{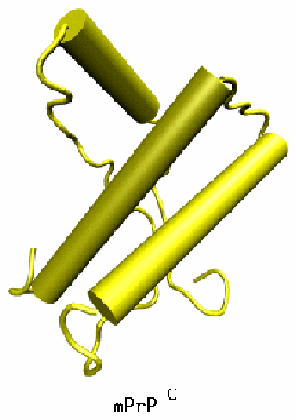}}\\
        \subfigure[]{
	  \vspace{-2.25cm}
	  \hspace{-2.1in}
        \label{fig:struct_mH23_eq300_91_mPrPCstar_2}
        \includegraphics[width=2.75in]{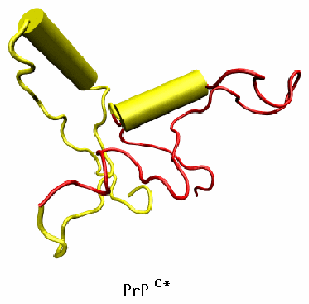}}
	\subfigure[]{
	  \vspace{-2.25cm}
	  \hspace{5.7cm}
        \label{fig:struct_mH23_Moil}
        \includegraphics[width=2.75in]{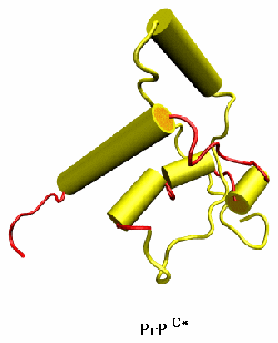}}
        \caption{}
\label{fig:snapshots_H23mPrP_SSon_Namd}
\end{sidewaysfigure}

The conformational transitions are correlated with an increase in the
angle between the axes of the two halves of H2 which changes from
20$^{o}$ to 90$^{o}$ (in the first 10 ns) followed by rapid
oscillations between these values for the remainder of the trajectory.
The transition is initiated in the second half of H2 where the
distances between (i,i+4) positions increase from 5 to 14 \AA\ in
about 10 ns. At longer time scales (t $\sim$ 60 ns) the distances
between (i,i+4) residues in the first half of H2 also increase from 5
to 13 \AA. These motions in H2 are correlated with fluctuations in H3,
where the distances between the first four (i,i+4) pairs of residues
in H3 and between positions 212 - 218 (with the exception of Cys214)
increase from 5 to 13 \AA\ in about 10 ns. Almost complete loss of
helical structure occurs towards the end of the trajectory
(Fig.(\ref{fig:snapshots_H23mPrP_SSon_Namd})). Thus, the conclusions
based on bioinformatic analysis are consistent with the results of MD
simulations.\\

\noindent
{\em Proposed structures for PrP$^{C*}$:} Our simulations
\cite{DimaPNAS04} and recent experiments
\cite{SpeareJBC03,ZieglerJBC03} strongly suggest that H1 is unlikely
to change conformation in the PrP$^{C} \rightarrow$ PrP$^{C*}$
transition. The most drastic change occurs in the second half of H2
and parts of H3. Based on the assumption that alterations in the
conformation of H2+H3 do not significantly affect the rest of the
protein, we have constructed a plausible ensemble of structures for
PrP$^{C*}$ (Fig.(\ref{fig:struct_mH23_eq300_91_mPrPCstar_2}) and
\ref{fig:struct_mH23_Moil}).  In PrP$^{C*}$(90-231) obtained from the
NAMD trajectories (Fig.(\ref{fig:struct_mH23_eq300_91_mPrPCstar_2}))
the helical content is $\sim$ 20\% (a lower bound), and in
PrP$^{C*}$(90-231) reached during the MOIL simulations
(Fig.\ref{fig:struct_mH23_Moil}) the helical content is $\sim$ 30\%
compared to 48\% in mPrP$^{C}$(90-231).

The overall characteristics of these structures are consistent with
those proposed by James and coworkers \cite{KuwataBiochem02}. It
remains to be seen if formation of PrP$^{C*}$, with fluctuating
regions in H2+H3, is required for oligomerization of PrP$^{C}$ i.e.,
if PrP$^{C*}$ is an on-pathway monomeric intermediate on the route to
fibrillization. We should emphasize that the conformation of the prion
protein in PrP$^{Sc}$ need not coincide with PrP$^{C*}$.\\

\begin{figure}[ht]
\includegraphics[width=4.5in]{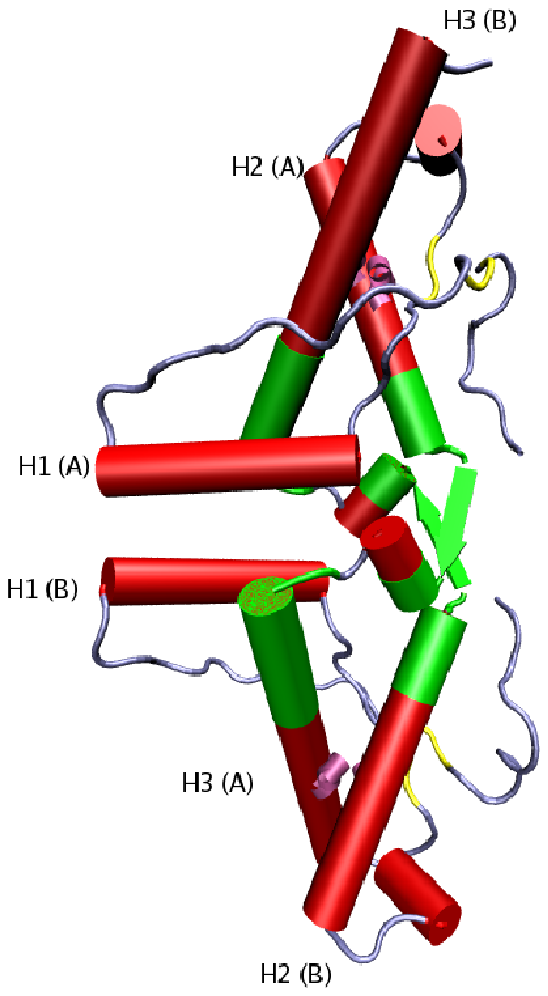}
\caption{\label{fig:1i4m_dimer}}
\end{figure}

\noindent
{\em Comparison of PrP$^{C*}$ structure with the Human prion protein
dimer:} In an important paper Knaus et al. \cite{KnausNSB01} announced
a 2 \AA crystal structure of the dimeric form of the human prion
protein (residues 90-231). The structure (Fig.(\ref{fig:1i4m_dimer}))
suggest that dimerization occurs by domain swap mechanisms in which H3
from one monomer packs against H2 from another. In fact, Eisenberg and
coworkers have suggested that domain swapping mechanism may be a
general route of amyloid fibril formation \cite{EisenbergNSB02}. The
electron density map seems to suggest structural fluctuations in the
residues 189-198 which coincides with the maximally frustrated region
predicted theoretically. The dimer interface is stabilized by residues
that are in H2 in the monomeric NMR structures. The header of the PDB
file of the monomeric structure of human PrP$^C$ indicates that H2
ends at residue 194 and H3 begins at 200. The domain swapped dimer
structure shows that residues 190-198 exist largely in a
$\beta$-strand conformation. The $\alpha \rightarrow \beta$ transition
minimizes frustration. An implication of the dimer structure is that
oligomerization occurs by domain swapping which in PrP$^C$ also might
implicate the disulfide bond between Cys residues at 179 and 214. The
role of the S-S bond in the PrP$^{Sc}$ formation remains
controversial. In our full-atom MD simulations of reduced mouse
PrP$^{C}$ (data not published), we found structures that closely
resemble the monomeric structure in this dimer. For example, H1
remains mostly intact, while H2 breaks into two smaller helices, one
running from its normal N-term end to position 187 and the other being
formed by the C-term end residues of the original H2 and residues from
the loop connecting it to H3. These findings suggest that the dimer
structure is a likely route to unfolding and self-assembly of
monomeric PrP$^{C}$.

Sequence pattern matches and long multiple molecular dynamics
simulations of helix 1 in mPrP$^{C}$ using two force fields show that
the stability of H1 is due to the formation of stabilizing internal
salt bridges. In view of the high propensity of $\alpha$-helix
observed in the isolated H1 in conjunction with supporting
experimental results \cite{LiuBiopolym99,SpeareJBC03,ZieglerJBC03} it
is clear that alterations in the conformation of H1 are unlikely in
the PrP$^{C} \rightarrow$ PrP$^{C*}$ transition.

The predicted tendency for the second half of H2 to be involved in the
formation of PrP$^{C*}$ is also consistent with the observation that a
number of mutations at 187 and 188 (H187R, T188R, T188K, and T188A)
are associated with various prion diseases. Based on our findings we
proposed that regions 186-190 and 214-226 must play a central role {\em
in the initial stages} that involve the PrP$^{C} \rightarrow$
PrP$^{C*}$ transition.  The large conformational change is likely to
be accompanied by stretching and rotation of the two halves of H2 and
by the unwinding of the N-terminal end of H3. The formation of the
domain swapped structure in the dimer structure of human PrP$^{C}$
\cite{KnausNSB01} might be facilitated by these large scale motions.\\

\noindent
{\bf Conclusions} 

\noindent
The development of methods to envision the structure of amyloid
fibrils has enabled us to obtain molecular insights into the assembly
process itself. Computational and experimental studies are beginning
to provide detailed information, at the residue level, about the
regions in a  given protein that harbor amyloidogenic tendencies. We
have harnessed these developments to propose tentative ideas on the
molecular basis of protein aggregation. These principles (or, more
precisely, rules of thumb) may be useful in the interpretation and design
of new experiments. 

Examination of the stable structures of oligomers and fibrils obtained
using experimental restraints and simulations show that these must be
stable conformations which maximize the inter-peptide interactions and
minimize electrostatic repulsions. Broadly, this is the only amyloid
self-organization principle (ASOP) that seems to be obeyed. From the
ASOP it follows that the formation of amyloid fibrils should indeed be
a generic property of almost all proteins and peptides under suitable
conditions. If this were the case then it is remarkable that in the
case of normal function aggregation is avoided most of the time. The
lack of preponderant protein aggregation may well be due to the
efficiency of cleaning mechanisms operating in the cell. This may
explain the lack of aggregation of PrP$^{C*}$ under most
circumstances. We conjecture that because of efficient degradation
processes only mild sequence constraints are needed to prevent
oligomer formation during the typical life cycle of newly
synthesized proteins.

From a biophysical perspective there are a number of open
problems. Are there common pathways involved in the self-assembly of
fibrils? Because of the paucity of the structural description of the
intermediates involved in an aggregation process a definitive answer
cannot be currently provided. The energy landscape perspective,
summarized briefly in Fig.(\ref{fig:Fig4_COSBreview}), suggests that
multiple scenarios for assembly must exist. Although the generic
nucleation and growth governs fibril formation the details can vary
considerably. The microscopic basis for the formation of distinct
strains in mammalian prions and in yeast prions remains a mystery. Are
these merely associated with the heterogeneous seeds or are there
unidentified mechanisms that lead to their growth? What factors may
determine the variations in the fibrillization kinetics for the wild
type and the mutants? A tentative proposal is that the kinetics of
polymerization is determined by the rate of production of {\bf
N$^\star$} (Fig. (\ref{fig:Fig4_COSBreview})) \cite{Regan00PNAS},
which in turn is controlled by barriers separating {\bf N} and {\bf
N$^\star$} \cite{Kelly02PNAS,DimaBiophysJ02}.  In this scenario the
stability of {\bf N} plays a secondary role. The generality of this
observation has not yet been established. Finally, how can one design
better therapeutic agents based on enhanced knowledge of the assembly
mechanism? Even in the case of sickle cell disease viable therapies
began to emerge only long after the biophysical aspects of gelation
were understood \cite{Eaton95Science}.

\newpage

{\bf Figure Captions}

Fig.\ref{fig:Fig4_COSBreview}. Schematic diagram of the two plausible
scenarios of fibrillization based on free energy landscape
perspective.  According to scenario I, the assembly competent state
{\bf N$^\star$} is metastable with respect to the monomeric native
state {\bf N} and is formed due to partial unfolding. In scenario II
{\bf N$^\star$} is formed upon structural conversion either of the
native state {\bf N} (as in prions) or directly from the unfolded
state {\bf U} (as in A$\beta$-amyloid peptides).  In both cases
proteins (or peptides) in {\bf N$^\star$} states must coalescence into
larger oligomers capable of growth into fibrils.

Fig.\ref{fig:Fig2_JMBTarus}. The putative dimer structures
corresponding to the \mbox{$\varphi$-dimer} (a) and
\mbox{$\varepsilon$-dimer} (b), respectively. The side chains at the
dimer interface are depicted explicitly. The positively and negatively
charged, polar, and hydrophobic residues are colored in blue, red,
purple, and green, respectively. The $\rm{C}_{\alpha}$ atoms of the
monomers A (left) and B (right) are colored in cyan and yellow,
respectively.

Fig.\ref{fig:Fig5_JMBTarus}. The Potential of Mean Force (PMF) is
plotted for two different relative orientations of the monomeric
peptide within the dimer. The PMF is computed as a function of the
surface separation, $\delta=\xi-\xi_{cont}$, along the distance
between the centers of mass (DCOMs) of the two monomers, where $\xi$
and $\xi_{cont}$ are the DCOMs of the two monomers when they are at an
arbitrary separation and in contact, respectively. The profile in blue
corresponds to the free energy surface computed using the
\mbox{$\varepsilon$-dimer} as the starting structure. The red curve is
similarly computed using the \mbox{$\varphi$-dimer} as the starting
structure. The difference between the two surfaces suggests that
hydrophobic interactions may be more essential to stabilization of the
dimer structure than electrostatic interactions.

Fig.\ref{fig:1qlx}. Cartoon representation of the structure of human
PrP$^{C}$ (PDB entry 1QLX). The three helices in the 90-231 ordered
region of PrP$^{C}$ are colored in red, while the short $\beta$-sheet
is in yellow. The two cysteine residues (179 and 214) involved in the
disulfide bond that connect H2 with H3 are indicated in bond
representation and colored in purple. The C-term end of H2 and the
N-term end of H3 which we believe to be implicated in the initial
stages of the $\alpha \rightarrow \beta$ transition are colored in
green. The figure was produced with packages VMD
\cite{HumphreyJMolGraph96} and PovRay (http://www.povray.org/).

Fig.\ref{fig:1lg4}. Cartoon representation of the structure of human
Doppel protein (Dpl) (PDB entry 1LG4). The three helices in the 24-152
ordered region of Dpl are colored in red. The four cysteine residues
(94, 108, 140 and 145) involved in the two disulfide bonds that
connect H2 with H3 and the loops preceding them are indicated in bond
representation and colored in purple. The figure was produced with
packages VMD \cite{HumphreyJMolGraph96} and PovRay
(http://www.povray.org/).

Fig.\ref{fig:snapshots_H23mPrP_SSon_Namd}. Schematic representation of
PrP$^{C} \rightarrow$ PrP$^{C*}$ transition, where the conformation
for PrP$^{C}$ is taken from the PDB file 1ag2 (yellow). The
conformations for PrP$^{C*}$ contain H1 from 1ag2 while the residues
encompassing H2+H3 are shown in a conformation (red) reached towards
the end of our MD simulations using the NAMD package
(\ref{fig:struct_mH23_eq300_91_mPrPCstar_2}) or the simulations using
the MOIL package (\ref{fig:struct_mH23_Moil}).  The schematic
PrP$^{C*}$ structures are representatives from ensembles of
fluctuating conformations.  In the representative PrP$^{C*}$ structure
obtained using NAMD simulations the H1 region, together with the
adjacent loops and the $\beta$-strands, and residues (205-212) from H3
retain their original conformations and are therefore depicted with
same color as in PrP$^{C}$. In the MOIL representative PrP$^{C*}$
structure the H1 region, together with the adjacent loops and the
$\beta$-strands, and residues (175-179), (184-188), (193,194) from H2
and residues (203-218) from H3 retain their original conformations and
are therefore depicted with same color as in PrP$^{C}$. The figures
are rotated such that the orientation of H1 is the same in all of
them.  The figures were produced with packages VMD
\cite{HumphreyJMolGraph96} and PovRay (http://www.povray.org/).

Fig.\ref{fig:1i4m_dimer}. Cartoon representation of the X-ray crystal
structure of the human PrP$^{C}$ dimer (PDB entry 1I4M). For each
chain, A and B, the three helices in the 90-231 ordered region of
PrP$^{C}$ are colored in red, while the short $\beta$-sheet is in
yellow. The two cysteine residues (179 and 214) involved in the
disulfide bond that connect H2 with H3 are indicated in bond
representation and colored in purple. The C-term end of H2 and the
N-term end of H3 which we believe to be implicated in the initial
stages of the $\alpha \rightarrow \beta$ transition are colored in
green. We notice that, in contrast to the monomeric PrP$^{C}$
structure from Fig.(\ref{fig:1qlx}), here this region is no longer
entirely helical, but contains a short stretch of $\beta$-strand
structure and a shorter helix as well. The figure was produced with
packages VMD \cite{HumphreyJMolGraph96} and PovRay
(http://www.povray.org/).

\[
\]

\[
\]

\[
\]

\[
\]

\[
\]

\end{document}